\documentclass[numberedappendix]{emulateapj} 

\usepackage{natbib}
\usepackage{hyperref}
\hypersetup{colorlinks,citecolor=Blue,linkcolor=Red,urlcolor=Blue}
\usepackage{amsmath}
\usepackage{empheq}
\usepackage{mathrsfs}
\usepackage[usenames,dvipsnames]{color}

\allowdisplaybreaks 

\begin{document}
 
\title{Generation of Highly Inclined Trans-Neptunian Objects by Planet Nine}  

\shorttitle{Make the Kuiper Belt Great Again!} 
\shortauthors{Batygin \& Brown} 

\author{Konstantin Batygin \& Michael E. Brown} 

\affil{Division of Geological and Planetary Sciences, California Institute of Technology, Pasadena, CA 91125} 

\email{kbatygin@gps.caltech.edu}
 
\newcommand{\Nice}{\textit{Nice}}

\begin{abstract} 
The trans-Neptunian region of the solar system exhibits an intricate dynamical structure, much of which can be explained by an instability-driven orbital history of the giant planets. However, the origins of a highly inclined, and in certain cases retrograde, population of trans-Neptunian objects remain elusive within the framework of this evolutionary picture. In this work, we show that the existence of a distant, Neptune-like planet that resides on an eccentric and mildly inclined orbit fully accounts for the anomalous component the trans-Neptunian orbital distribution. Adopting the same parameters for Planet Nine as those previously invoked to explain the clustering of distant Kuiper belt orbits in physical space, we carry out a series of numerical experiments which elucidate the physical process though which highly inclined Kuiper belt objects with semi-major axes smaller than $a<100\,$AU are generated. The identified dynamical pathway demonstrates that enigmatic members of the Kuiper belt such as Drac and Niku are derived from the extended scattered disk of the solar system.
\end{abstract} 

\maketitle

\section{Introduction} \label{sect1}

The detection and observational characterization of the Kuiper belt have caused a qualitative shift in our understanding of the solar system's post-nebular evolution. The gradual unveiling of the trans-Neptunian region's orbital distribution has led to the replacement of a largely static solar system formation scenario \citep{Cameron1988,Lissauer1993} with a dynamic picture, wherein long-range planetary migration is facilitated by the onset of a transient dynamical instability \citep{Tsiganis2005,Morbidelli2008}. This new class of instability-driven models, collectively known as the \Nice\ model, has been remarkably successful in explaining a number of perplexing features within the solar system.

The \Nice\ model's list of accolades begins with the reproduction of the solar system's planetary architecture itself \citep{Tsiganis2005,BatBro2010}. The \Nice\ model further explains the origins of Jovian and Neptunian co-orbital (Trojan) populations \citep{Morbidelli2005,Nesvorny2009}, while simultaneously providing a natural trigger for the Lunar late heavy bombardment \citep{Gomes2005,Levison2011}. Finally, the \Nice\ model successfully accounts for the dynamically excited orbital distribution of the resonant, hot classical, as well as scattered disk sub-populations of the Kuiper belt\footnote{The so-called cold classical component of the Kuiper belt likely formed \textit{in-situ}, and the \Nice\ model largely preserves its primordial unperturbed state \citep{Batetal2011,Nesvorn2015a}.} \citep{Levison2008}. Despite these successes, however, the \Nice\ model fails to explain a notable, highly inclined subset of trans-Neptunian objects (TNOs), leaving the physical mechanism responsible for their nearly orthogonal and retrograde orbits elusive. The origin of this remarkable group of small bodies is the primary focus of this paper.

Dynamical emplacement of icy debris into the Kuiper belt during the epoch of Neptune's migration can yield inclinations as large as $i\sim40-60\deg$ (the exact value depends on the details of Neptune's assumed evolution; \citealt{Nesvorn2015b}). Although substantial, such inclination are dwarfed by the nearly-perpendicular orbits of the TNOs Drac (2008\,KV$_{42}$; $a=41\,\mathrm{AU},e=0.5,i=103\deg$; \citealt{Gladman2009}) and Niku (2011\,KT$_{19}$; $a=36\,\mathrm{AU},e=0.3,i=110\deg$; \citealt{Chen2016}). Even more dramatically, the object 2016\,NM$_{56}$ ($a=74\,\mathrm{AU},e=0.9,i=144\deg$) occupies a retrograde orbit that is relatively close to the plane of the solar system. Simply put, there exists no physical mechanism to produce such inclinations within the framework of the \Nice\ model. How then, are such highly inclined orbits generated? 

\begin{figure*}[t]
\centering
\includegraphics[width=1\textwidth]{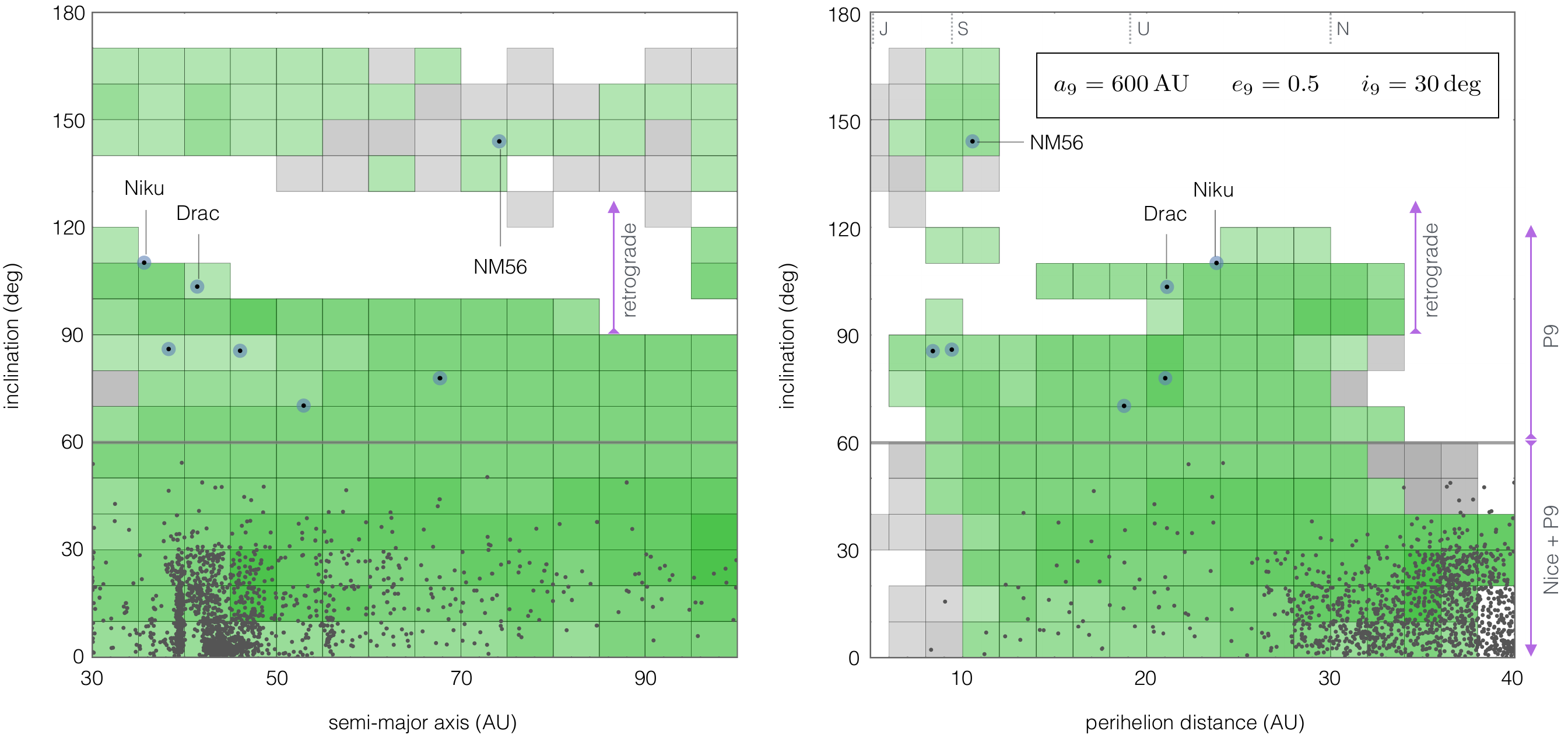}
\caption{Orbital distribution of the trans-Neptunian region. The current observational census of all TNOs with $a\in(30,100)\,$AU is shown. Regular objects with $i<60\deg$ are depicted with small black dots, and the anomalous objects with $i>60\deg$ are emphasized with blue circles. The three retrograde orbits of Drac, Niku, and 2016\,NM$_{56}$ are labeled explicitly. The theoretically computed orbital distribution is shown as a density histogram that underlies the observational data. Gray and green colors denote orbital paths traced by particles with dynamical lifetimes in the range of $3-4\,$Gyr and in exceess of $4\,$Gyr respectively. Simultaneously, transparency is used as a proxy for the amount of time spent by particles in a given region of orbital element space, with solid color corresponding to a higher probability of dynamical emplacement. Clearly, the entire collection of $i>60\deg$ objects, including those occupying retrograde orbits, is adequately explained by the existence of Planet Nine.}
\label{f1}
\end{figure*}

In a recent study \citep{BatBro2016}, we proposed that the physical alignment of TNO orbits with semi-major axes greater than $a\gtrsim150-250\,$AU and perihelion distance beyond $q\gtrsim30\,$AU can be explained by the existence of an additional Neptune-like planet, which resides on a distant, eccentric, and moderately inclined orbit. In this work, we refer to this object as ``Planet Nine." Numerical simulations reported in \citet{BatBro2016} and \citet{BroBat2016} suggest that this body has a mass of $m_9\sim10\,m_{\oplus}$, perihelion distance of $q_9\sim250\,$AU, and a semi-major axis of $a_9\sim 400-700\,$AU. 

Gravitational torques exerted by Planet Nine onto the small bodies it shepherds, manifest in an extensive web of mean-motion resonances, which maintain a rough co-linearity of distant TNO orbits over multi-Gyr timescales. Importantly, however, not all trajectories remain physically confined: Kozai-Lidov type interactions \citep{Lidov1962,Kozai1962} driven by Planet Nine can dramatically modulate the eccentricities and inclinations of distant KBOs, thereby reproducing a population of highly inclined ($i>40\deg$), large semi-major axis ($a>100\,$AU) Centaurs \citep{Gomes2015}. With this notion in mind, \textit{here we investigate the possibility that Planet Nine can self-consistently explain the unusual inclinations of Drac, Niku, 2016\,NM$_{56}$, as well as every other member of the trans-Neptunian population with inclinations greater than $i>60\deg$ and semi-major axes below} $a<100\,$AU i.e. outside of the range of possibilities of the \Nice\ model and outside Planet Nine's region of direct gravitational influence. 

The paper is organized as follows. We describe the details of our numerical experiments in section \ref{sect2}, and present the results in section \ref{sect3}. We summarize and discuss the implications of our findings in section \ref{sect4}.

\section{Numerical Experiments} \label{sect2}
In order to investigate the dynamical behavior of small bodies under the influence of Jupiter, Saturn, Uranus, Neptune, as well as Planet Nine, we have carried out a series of numerical $N$-body experiments. The performed simulations were evolutionary in nature: a synthetic solar system, with initial conditions corresponding to the final stages of the \Nice\ model \citep{Levison2008,Batetal2011,Nesvorn2015b}, was evolved forward in time for 4 Gyr. Following the numerical calculations reported in \citet{BatBro2016} and \citet{BroBat2016}, the starting configuration of the Kuiper belt was represented by a disk of 3,200 test particles, uniformly distributed across the semi-major axes range $a\in(150,550)\,$AU and perihelion distance $q\in(30,50)\,$AU. The initial inclinations were drawn from a half-normal distribution with a standard deviation of $\sigma_i=15\deg$, while the remaining orbital angles (namely, argument of perihelion, longitude of ascending node, and mean anomaly) were assigned random values between $0$ and $360\deg$. 

In contrast with our previous models, here we did not mimic the orbit-averaged gravitational field of Jupiter, Saturn, and Uranus with an enhanced quadrupolar component of the Sun's potential. Instead, we modeled the planets in a direct $N-$body fashion, fully resolving their Keplerian motion. For our nominal run, the known giant planets were placed on their current orbits, while Planet Nine was chosen to have a mass $m_9=10\,m_{\oplus}$ and initialized on an orbit with $a_9=600\,$AU, $e_9=0.5\,$, $i_9=30\deg$, and $\omega_9=150\deg$. We note that although this choice of parameters is marginally different from the $a_9=700\,$AU, $e_9=0.6\,$AU, $i_9=30\deg$ Planet Nine considered in \citet{BatBro2010}, it generates a synthetic Kuiper belt that provides an optimal match to the existing observations \citep{BroBat2016}. A detailed analysis of the synthetic distant Kuiper belt sculpted in this simulation is presented in our companion paper \citep{BroBat2016b}.

The calculations were carried out using the \texttt{mercury6} gravitational dynamics software package \citep{Chambers1999}. A hybrid Wisdom-Holman/Bulirsch-Stoer algorithm \citep{WisdomHolman1992,Press1992} was employed for all simulations, adopting a time-step of $\tau=300\,$days. The presence of the terrestrial planets was ignored, and any object that attained a radial distance smaller than $r<5\,$AU or larger than $r>10,000\,$AU was removed from the simulation. All orbital elements were measured with respect to Jupiter's orbital plane, which precesses by a few degrees over the lifetime of the solar system due to the gravitational influence of Planet Nine \citep{Bailey2016,Gomes2016}.

In addition to our nominal case, we carried out a series of simulations with lower particle count, sampling the favorable locus of parameter space identified in \citet{BroBat2016}. To this end, we found that the low-inclination component of the distant Kuiper belt is far more sensitive to the specific orbit of Planet Nine than the high-inclination component of the sculpted test particle population (which is the primary focus of this paper). As a result, here we restrict ourselves to presenting only the results from our nominal calculation, keeping in mind that they can be deemed representative for any reasonable choice of Planet Nine's parameters. 

\section{Results} \label{sect3}
The current observational census of multi-opposition TNOs with semi-major axes in the range $a\in(30,100)\,$AU is presented in Figure (\ref{f1}). Objects with inclinations greater than $i>60\deg$ are emphasized, as they represent the anomalous component of the orbital distribution. The left and right panels show inclination as a function of semi-major axes and perihelion distance respectively.

Recalling that the entire small body population is initialized with $a>150\,$AU in the simulations, we have analyzed the dynamical evolution of the test particles with an eye towards identifying objects that veer into the region depicted in Figure (\ref{f1}). Owing to chaotic variations of the orbits, instances where objects enter this domain are in fact quite common. However, these excursions are often followed by ejection from the solar system. As a result, it is sensible to focus only on objects whose dynamical lifetime is comparable to the age of the sun.

The simulated orbital distribution of long-term stable bodies is shown as a density histogram that underlies the observational data in Figure (\ref{f1}). Green squares correspond to the orbital distribution traced out by objects that are stable over the full 4 Gyr integration period, while the gray squares represent bodies with dynamical lifetimes between 3 and 4 Gyr. Transparency of the color is used as a logarithmic proxy for the amount of time the particles spend in a given box, with solid colors corresponding to regions of higher visitation probability.

Clearly, the entire observational data set of TNOs with $i>60$, including the retrograde orbits of Drac, Niku, and 2016\,NM$_{56}$, is well explained by the simulation results. Simultaneously, it is noteworthy that the theoretical inclination distribution is not uniform. Instead, it is comprised of two components: one that extends from $i=0$ to $i\sim110\deg$ and a second, somewhat less densely populated component that is centered around $i\sim150\deg$. These two constituents also differ in their characteristic perihelion distances, with the lower inclination part extending from $q\sim5\,$AU to $q\sim35\,$AU, and the higher inclination part characterized by substantially lower values of $q\sim10\,$AU.

\begin{figure}[t]
\centering
\includegraphics[width=1\columnwidth]{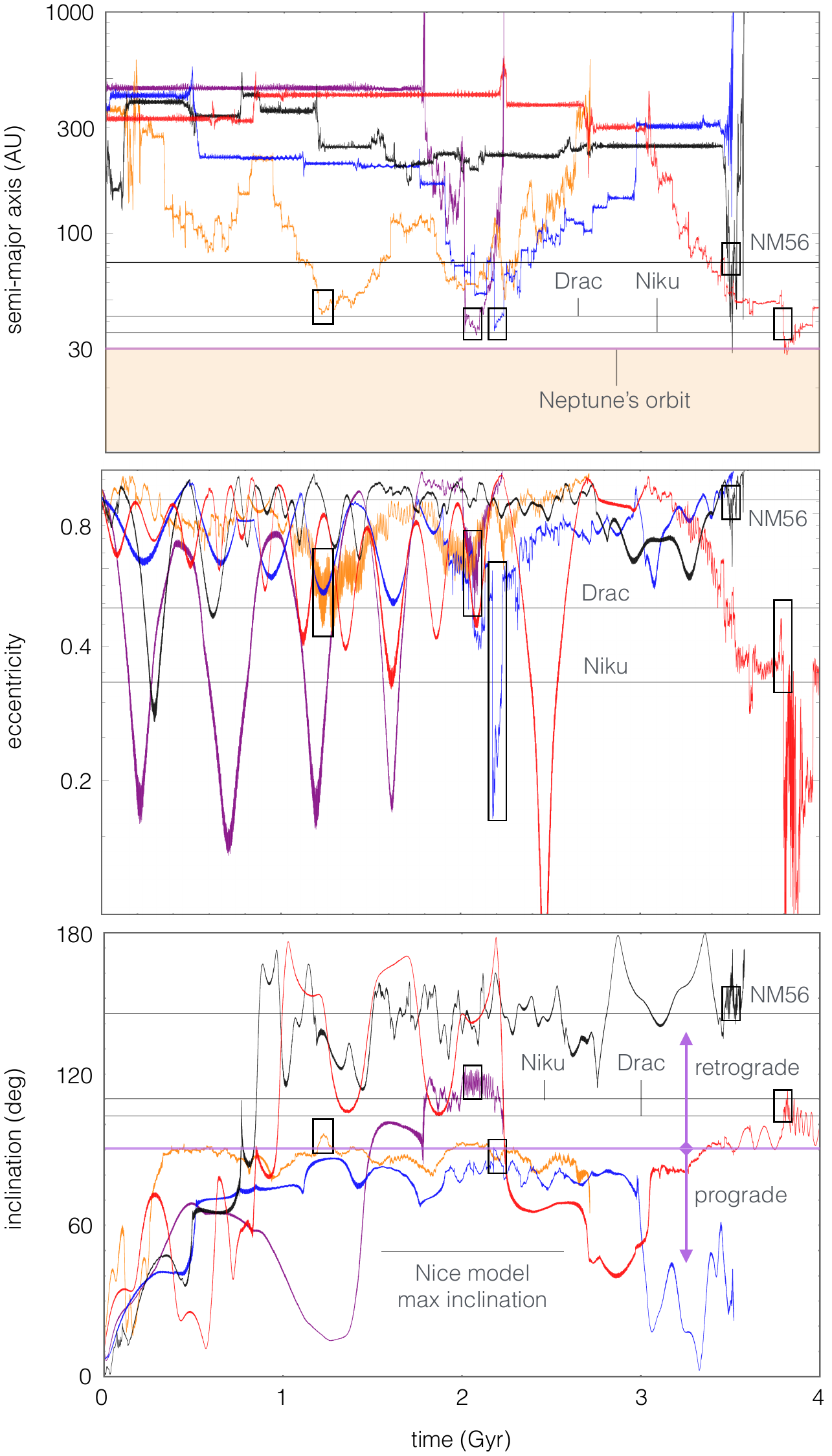}
\caption{Time-series of a subset of simulated particles that attain highly inclined orbits with $a<100\,$AU within the span of the integration. The colors hold no physical meaning and simply label different bodies. The orbital parameters of the three retrograde members of the observational census are shown with horizontal lines on each panel, and points in time when simulated objects attain orbits that are close to their observed counterparts are emphasized with black squares. The two objects with longest dynamical lifetimes are particularly notable, as they reproduce the orbits of Drac, Niku, and 2016\,NM$_{56}$ almost exactly. Additionally, the object depicted in blue exemplifies a reversal of the dynamical pathway for generation of Drac-type orbits from the distant Kuiper belt, as it reacquires a typical scattered disk orbit towards the end of the simulation.}
\label{f2}
\end{figure}

What is the physical mechanism through which the simulated particles acquire these unusual orbits? Our calculations indicate that a sequential combination of Kozai-Lidov cycles driven by Planet Nine and close encounters with Neptune plays a dominant role. First, Kozai-Lidov interactions induce large-scale oscillations in the inclinations and eccentricities of particles with initially large semi-major axes. Thus, the perihelion distance of a typical low-inclination scattered disk object can grow to larger values, only to recede back down to hug the orbit of Neptune at a much higher inclination. Then, close encounters that inevitably ensue, facilitate a stochastic evolution of the semi-major axis, occasionally reducing it below $a<100\,$AU.

To illustrate this process, Figure (\ref{f2}) shows the time-series of a subset of simulated objects that come to resemble Drac and Niku, or 2016\,NM$_{56}$ at one point in time. The top, middle, and bottom panels depict semi-major axes, eccentricities, and inclinations as functions of time respectively, while the boxes indicate the phases of dynamical evolution where simulated particles attain orbits that are close to those of the observed retrograde objects. As can be deduced by examining the individual paths of the particles in detail, the aforementioned qualitative picture holds, although the orbits generally exhibit chaotic motion. Among the exemplified evolutions (which represent only a small subset of the entire simulation suite), the two objects with the longest dynamical lifetimes (shown in black and red) are particularly notable. Starting out on low-inclination, high-eccenticity orbits with semi-major axes beyond $a>300\,$AU, these bodies evolve to attain orbital states that are almost exact replicas of Drac, Niku, and 2016\,NM$_{56}$. 

Conventional $N-$body simulations of TNOs in this class (e.g. \citealt{Gladman2009,Chen2016}; see also \citealt{Dones1996,2007Icar..190..224D}) yield strictly unstable orbits with dynamical lifetimes of order $\sim10\,$Myr$-1\,$Gyr. Although future ejection of these bodies is indeed a distinct possiblity, our simulations suggest that the existence of Planet Nine can potentially prolong their lifetimes, through a reversal of their delivery process. An example of such behavior is demonstrated by the orbit depicted in blue in Figure (\ref{f2}) - after reaching a semi-major axis of $a\sim35\,$AU at $t\sim2.2\,$Gyr, the object's perihelion distance and semi-major axes increase to values comparable to their starting conditions. The concurrent decrease in the inclination means that by the time the object escapes from the solar system at $t\sim3.5\,$Gyr, it bears semblance to a regular member of the distant Kuiper belt. Therefore, our simulations not only reveal that Drac, Niku and other TNOs occupying highly inclined orbits are sourced from the extended scattered disk, they also point to an evolutionary future where some of these bodies will once again return to more conventional, low-inclination orbits. 

\section{Discussion} \label{sect4}
In this work, we have carried out a sequence of numerical experiments, with an eye towards exploring the dynamical effects of Planet Nine onto the $a<100\,$AU portion of the Kuiper belt. Adopting the same orbital parameters for Planet Nine as those required to generate physical confinement among distant ($a>150\,$AU) TNO orbits \citep{BatBro2016,BroBat2016}, we have shown that Planet Nine is capable of explaining the full range of inclinations observed in the Kuiper belt. The origin of these unusual objects had remained elusive until now \citep{Gladman2009,Nesvorn2015b}, and the existence of Planet Nine provides a resolution for this puzzle.

While the orbital domain currently occupied by the highly inclined sub-population of the Kuiper belt lies outside of Planet Nine's direct gravitational reach, our calculations suggest that these bodies had substantially larger semi-major axes in the past. Specifically, the simulations reveal a dynamical pathway wherein long period TNOs undergo Kozai-Lidov oscillations facilitated by Planet Nine, and subsequently scatter inwards due to close encounters with Neptune (as well as other known giant planets). This dynamical pathway is time-reversible, and the numerical experiments reveal examples of trajectories that originate within the extended scattered disk, proceed to become nearly orthogonal members of the classical Kuiper belt, and subsequently reacquire long orbital periods and low inclinations. This finding places large semi-major axis Centaurs \citep{Gomes2015} and retrograde Kuiper belt objects such as Drac and Niku into the same evolutionary context.

Although our model explains extant data adequately, it can be tested with further observations. To this end, we note that the simulations predict a rather specific orbital distribution in $a-q-i$ space. The explicit non-uniformity of this theoretical expectation renders our model readily falsifiable: detection of bodies within the empty region can constitute significant evidence against the dynamical mechanism described herein (although we caution that the empty region diminishes somewhat for simulations with smaller values of $a_9$, e.g. 500\,AU). We further note that there exist high inclination Centraurs with $a<30\,$AU in the observational data set. Here we have chosen to ignore this component of the data, because our model does not have sufficient resolution in terms of particle count to adequately model the inter-planetary region. However, we speculate that the evolutionary histories of these objects may be connected to the high-inclination bodies with $a>30\,$AU, and modeling their generation presents an interesting avenue for future research.

In concluding remarks, we wish to draw attention to the recent proposition of \citet{Chen2016}, who noted that all Centaurs and TNOs with $a<100\,$AU, $q>10\,$AU, and $i>60\deg$ appear to occupy a common plane. Our simulations do not show the existence of such a plane, and predict that future observations will reveal objects that do not correspond to the pattern pointed out by \citet{Chen2016}. Importantly, the latest addition to the observational census of trans-Neptunian objects, 2016\,NM$_{56}$, conforms to the aforementioned selection criteria, but does not lie in the same plane as the other objects in the group. This supports the notion that the apparent clustering of ascending node is not statistically significant. Future observations will continue to test the theoretical expectation outlined by our model. 
\\
\\
\textbf{Acknowledgments}  \\ 
We are thankful to Ira Flatow and Chris Spalding for inspirational conversations.


\begin{thebibliography} 


\bibitem[Bailey et al.(2016)]{Bailey2016} Bailey, E., Batygin, K., \& Brown, M.~E.\ 2016, arXiv:1607.03963 

\bibitem[Batygin \& Brown(2010)]{BatBro2010} Batygin, K., \& Brown, M.~E.\ 2010, \apj, 716, 1323 

\bibitem[Batygin et al.(2011)]{Batetal2011} Batygin, K., Brown, M.~E., \& Fraser, W.~C.\ 2011, \apj, 738, 13 

\bibitem[Batygin \& Brown(2016)]{BatBro2016} Batygin, K., \& Brown, M.~E.\ 2016, \aj, 151, 22 

\bibitem[Brown \& Batygin(2016a)]{BroBat2016} Brown, M.~E., \& Batygin, K.\ 2016a, \apjl, 824, L23

\bibitem[Brown \& Batygin(2016b)]{BroBat2016b} Brown, M.~E., \& Batygin, K.\ 2016b, in prep.




\bibitem[Cameron(1988)]{Cameron1988} Cameron, A.~G.~W.\ 1988, \araa, 26, 441 

\bibitem[Chambers(1999)]{Chambers1999} Chambers, J.~E.\ 1999, \mnras, 304, 793 

\bibitem[Chen et al.(2016)]{Chen2016} Chen, Y.-T., Lin, H.~W., Holman, M.~J., et al.\ 2016, \apjl, 827, L24 



\bibitem[Di Sisto \& Brunini(2007)]{2007Icar..190..224D} Di Sisto, R.~P., \& Brunini, A.\ 2007, Icarus, 190, 224 


\bibitem[Dones et al.(1996)]{Dones1996} Dones, L., Levison, H.~F., \& Duncan, M.\ 1996, Completing the Inventory of the Solar System, 107, 233 






\bibitem[Gladman et al.(2009)]{Gladman2009} Gladman, B., Kavelaars, J., Petit, J.-M., et al.\ 2009, \apjl, 697, L91 

\bibitem[Gomes et al.(2005)]{Gomes2005} Gomes, R., Levison, H.~F., Tsiganis, K., \& Morbidelli, A.\ 2005, \nat, 435, 466 

\bibitem[Gomes et al.(2015)]{Gomes2015} Gomes, R.~S., Soares, J.~S., \& Brasser, R.\ 2015, Icarus, 258, 37 

\bibitem[Gomes et al.(2016)]{Gomes2016} Gomes, R., Deienno, R., \& Morbidelli, A.\ 2016, arXiv:1607.05111 






\bibitem[Kozai(1962)]{Kozai1962} Kozai, Y.\ 1962, \aj, 67, 591 



\bibitem[Levison et al.(2008)]{Levison2008} Levison, H.~F., Morbidelli, A., Van Laerhoven, C., Gomes, R., \& Tsiganis, K.\ 2008, Icarus, 196, 258

\bibitem[Levison et al.(2011)]{Levison2011} Levison, H.~F., Morbidelli, A., Tsiganis, K., Nesvorn{\'y}, D., \& Gomes, R.\ 2011, \aj, 142, 152 

\bibitem[Lidov(1962)]{Lidov1962} Lidov, M.~L.\ 1962, \planss, 9, 719 


\bibitem[Lissauer(1993)]{Lissauer1993} Lissauer, J.~J.\ 1993, \araa, 31, 129 


\bibitem[Morbidelli et al.(2005)]{Morbidelli2005} Morbidelli, A., Levison, H.~F., Tsiganis, K., \& Gomes, R.\ 2005, \nat, 435, 462 

\bibitem[Morbidelli et al.(2008)]{Morbidelli2008} Morbidelli, A., Levison, H.~F., \& Gomes, R.\ 2008, The Solar System Beyond Neptune, 275 


\bibitem[Nesvorn{\'y} \& Vokrouhlick{\'y}(2009)]{Nesvorny2009} Nesvorn{\'y}, D., \& Vokrouhlick{\'y}, D.\ 2009, \aj, 137, 5003 

\bibitem[Nesvorn{\'y}(2015a)]{Nesvorn2015a} Nesvorn{\'y}, D.\ 2015a, \aj, 150, 68 
\bibitem[Nesvorn{\'y}(2015b)]{Nesvorn2015b} Nesvorn{\'y}, D.\ 2015b, \aj, 150, 73 




\bibitem[Press et al.(1992)]{Press1992} Press, W.~H., Teukolsky, S.~A., Vetterling, W.~T., \& Flannery, B.~P.\ 1992, Numerical recipes in FORTRAN. The art of scientific computing, Cambridge: University Press, 2nd ed. 





\bibitem[Tsiganis et al.(2005)]{Tsiganis2005} Tsiganis, K., Gomes, R., Morbidelli, A., \& Levison, H.~F.\ 2005, \nat, 435, 459 





\bibitem[Wisdom \& Holman(1992)]{WisdomHolman1992} Wisdom, J., \& Holman, M.\ 1992, \aj, 104, 2022 






\end{thebibliography}
\end{document}